\documentclass{amsart}

\newtheorem{theorem}{Theorem}[section]

\theoremstyle{definition}

\theoremstyle{remark}

\newcommand{\cA}{{\mathcal A}}
\newcommand{\cB}{{\mathcal B}}
\newcommand{\C}{{\mathcal C}}

\newcommand{\cH}{{\mathcal H}}

\newcommand{\cK}{{\mathcal K}}

\newcommand{\cS}{{\mathcal S}}

\newcommand{\bC}{{\mathbb{C}}}
\newsymbol\boxtimes 1202
\newcommand{\Cs}{{{$\hbox{\bf C}^*$}}}
\numberwithin{equation}{section}

\newcommand{\jed}{{\mathbb{I}}}
\begin{document}

\setlength{\baselineskip}{2\baselineskip}

\begin{titlepage}
%\title[$k$-decomposability]{On $k$-decomposability of positive maps}
\begin{center}
\vspace*{15mm}
{\LARGE ON POSITIVE MAPS, ENTANGLEMENT AND QUANTIZATION%\footnote{
%W.A.M. was supported }
}\\
\vspace{2cm}
\textsc{{\large W{\l}adys{\l}aw A. Majewski}\\
Institute of Theoretical Physics and Astrophysics\\
Gda{\'n}sk University\\
Wita Stwosza~57\\
80-952 Gda{\'n}sk, Poland}\\
\textit{E-mail address:} \texttt{fizwam@univ.gda.pl}\\
\end{center}
%\end{titlepage}
\vspace*{1cm}
%\noindent
\textsc{Abstract.}
We  outline the scheme for quantization of classical Banach space results
associated with some prototypes of dynamical maps and describe the 
quantization of correlations as well. A relation between these two areas is discussed.

\vspace{1cm} {\bf Mathematical Subject Classification}: Primary: 46L53, 46L60:
Secondary: 46L45, 46L30

\vspace{0.5cm}
 
\end{titlepage}

\newpage
\section{INTRODUCTION}
The aim of this paper is to bring together two areas, theory of positive maps on 
\Cs-algebras and theory of entanglement considered as a peculiar feature of non-commutative 
Radon measures. Both topics are at the heart of quantum theory, thus in particular, 
in the foundations of quantum information theory. It will be shown
that such pure quantum features as: peculiar behaviour of positive maps,
quantum correlations, entanglement, and quantum stochastic dynamics, can be easily obtained
within the framework of the \Cs-algebraic approach to Quantum Mechanics.
This approach sheds new light on entanglement and quantum features
of correlations of non-commutative systems.
In particular, some emphasis will be put on evolution of entanglement.

The paper is organized as follows.
Section 2 provides sufficient preparation for the concept of ``quantization'' 
of classical results related to 
prototypes of dynamical maps. Section 3 is concerned with entanglement
and the coefficient of quantum correlations. The latter is again 
an example of ``quantization'' of a classical concept.
The last section contains a brief discussion of applications of the presented results
to the description of quantum dynamical systems.
We will discuss the evolution of entanglement for some selected
models as well as relations between classification of positive maps
and measures of entanglement.

\section{POSITIVE MAPS}

In this section we compile some basic facts on the theory of positive maps on \Cs-algebras.
To begin with, let $\cA$ and $\cB$ be \Cs-algebras (with unit), $\cA_h = \{ a \in \cA;
a = a^* \}$, $\cA^+ = \{ a \in \cA_h; a \ge 0 \}$ - the set of all positive elements in 
$\cA$, and $\cS(\cA)$ the set of all states on $\cA$.
In particular
$$ (\cA_h, \cA^+)\quad is \quad an \quad ordered \quad Banach \quad space. $$
We say that a linear map $\alpha : \cA \to \cB$ is positive if $\alpha(\cA^+) \subset \cB^+$.

The theory of positive maps on non-commutative algebras can be viewed as a 
jig-saw-puzzle with 
pieces whose exact form is not well known. Therefore, as we address this paper to 
a readership interested in quantum mechanics and quantum information theory, 
we will focus our attention on ``quantization'' procedure of 
some classical (Banach space) results in order to  
facilitate access to some main problems of that theory.

We begin with the classical Banach-Stone result (\cite{B}, \cite{S})
{\it  if a unital linear map
$T: C(X) \to C(Y)$, where $X,Y$ are compact Hausdorff spaces, is either isometric or an 
order-isomorphism then it is also an algebraic isomorphism.} 
Thus, even in the Banach space setting, the order and algebraic structures are strongly related.
The Banach-Stone theorem has the following
non-commutative generalization (Kadison, \cite{K}): 
{\it a unital isometric or order isomorphic linear map $\alpha : \cA_h \to \cB_h$ must 
preserve the Jordan product ( $(a,b) \mapsto {1/2}(ab + ba)$)}.
In other words, this result indicates the role of a specific algebraic structure - 
the Jordan structure - in operator
algebras and that remark will be frequently used throughout the paper.
Moreover, such a result makes it legitimate to study and to classify \Cs-algebras $\cA$ 
by a detailed 
analysis of ordered Banach spaces $\cA_h$. However, this is {\it a very difficult task}.
In particular, it was soon realized that one of the basic problems is 
the answer to the following question: which compact convex sets can arise as the state spaces
of unital $C^*$-algebras (again a very difficult task!).

\smallskip

To describe the next result we need some preliminaries.
Let $(\Omega, \mu)$ be a measure space. Here and subsequently, $\mu$ stands for a 
probability measure.
The triple (semigroup $\{S_t\}$, $\Omega$, $\mu$) 
will denote the classical dynamical system where 
$ S_t : \Omega \to \Omega$ is a one parameter family of measure preserving maps.
The phase functions $f : \Omega \to \bC $  evolve according to the Koopman operators
$$ V_tf(\omega)=f(S_t\omega) \ , \ \ \omega\in\Omega \,. $$ 
It is known that
the Koopman operators $V_t$ are isometries on the Banach space 
$L^p=L^p(\Omega, \mu)$,  $p\ge 1$ of $p$-integrable functions and
unitary operators when  restricted to the Hilbert space $L^2$ and  the
transformations
$S_t$   are automorphisms. 
The relation of the point dynamics 
with the Koopman operators is clarified
by asking the question: {\it what 
types of isometries on   $L^p$
spaces are implementable by point transformations?} For $L^p$ spaces
$p\not=2$, all isometries induce underlying point transformations, i.e.
if $||Vf|| = ||f||$ for all $f \in L^p$, then $V$ is given by an underlying measurable 
point transformation $S$ and a certain function $h$ according to 
$(Vf)(x) = h(x)f(Sx)$.
 Such theorems on the
implementability of isometries on   $L^p$ spaces, $p\neq 2$, are known as
{\it Banach-Lamperti theorems } \cite{B}, \cite{La}.
They are of great importance for the Misra-Prigogine-Courbage
theory \cite{MPC} which is trying
{\it  to reconcile irreversible phenomena
with the basic dynamical laws}.

Again, one may ``quantize'' Banach-Lamperti theorems \cite{Ye}, see also \cite{AMS}.
To this end one should use the so called 
non-commutative (quantum) $L_p$-spaces. Namely, using  the ``quantized'' measure theory,
 let  $\{ \cA, \varphi \}$ be a von Neumann algebra with faithful
normal trace and let $L_p(\cA, \varphi)$, $p \ge 1$, be the corresponding
quantum $L_p$-space, i.e. a Banach space of operators which is closed under an appropriate norm. 
Assume that $T: L_p(\cA, \varphi) \to
L_p(\cA, \varphi)$  is a linear map.
{\it Then $T$ is an $L_p$-isometry if and only if }
$$T(x) = W B J(x), \quad x \in  L_p(\cA, \varphi) \cap \cA
$$
where $W \in \cA$ is a partial isometry, $B$ a selfadjoint operator
affiliated with $\cA$, $J$ a normal
Jordan isomorphism mapping $\cA$ into a weakly closed
$^*$-subalgebra of $\cA$ such that
$ W^*W = J(\jed) = supp(B)$ and $B$ commutes strongly with $ J(\cA)$. Again, we can see the 
importance of the Jordan structure.

\smallskip

The third example we wish to recall is associated with a very strong notion of positivity:
the so called complete positivity (CP).
Namely, a linear map $\tau : \cA \to \cB$ is CP iff
$$\tau_n : M_n(\cA) \to M_n(\cB); [a_{ij} \mapsto [\tau(a_{ij})]$$
is positive for all n.

To explain the basic motivation for that concept we need the following notion: 
{\it an operator state
of $\cA$ on a Hilbert space $\cK$ is a CP map $\tau : \cA \to \cB(\cK)$}. 
Having that concept we can recall the Stinespring result, \cite{Sti}, 
which is the generalization of GNS construction and which was the starting point for
a general interest in the concept of complete positivity.

{\it For operator state $\tau$ there is a Hilbert space $\cH$, 
a $^*$-representation $\pi : \cA \to
\cB(\cH)$ and a partial isometry $V : \cK \to \cH$ for which}
$$\tau(a) = V^* \pi(a) V.$$

Following the quantization ``route'', it was shown
\begin{itemize}
\item (Choi, \cite{Ch}) if $\tau : \cA \to \cB$ is a CP order isomorphism then it is
a $^*$-isomorphism. This can be considered as a final ``quantization'' of the Banach-Stone theorem.
\item (Arverson, \cite{Ar}) the Hahn-Banach theorem and its order-theoretical
version (due to Krein) has a nice generalization for non-commutative structures
 in terms of CP maps:  {\it Let ${\mathcal N}$ be a closed
self-adjoint subspace of \Cs algebra $\cA$ containing the identity and let
$\tau: {\mathcal N} \to \cB(\cH)$ be a CP map. Then $\tau$ possesses
an extension to a CP map $\tilde{\tau} : \cA \to \cB(\cH)$.}
\end{itemize}

It is worth pointing out that plain positivity is not enough for these  
generalizations. Moreover, Arverson's extension theorem is the basis 
of the CP ideology in open system theory.

\smallskip

Up to now we considered linear positive maps 
on an algebra without entering into the (possible) complexity of 
the underlying algebra. The situation changes when one is dealing 
with composed systems (for example in the framework of 
open system theory). Namely, 
there is a need to use the tensor product structure. In particular, 
again, we wish to consider positive maps but now defined on the tensor product
of two \Cs-algebras, 
$\tau : \cA \otimes \cB \to \cA \otimes \cB$.
But now the question of order is much more complicated. Namely, 
there are various cones determining the order structure in the tensor product of algebras (cf. \cite{W})
$$ \C_{inj} \equiv (\cA \otimes \cB)^+ \supseteq, ..., 
\supseteq \C_{\beta} \supseteq,..., \supseteq \C_{pro}\equiv
conv(\cA^+\otimes\cB^+)$$
and correspondingly in terms of states (cf \cite{MM})
$$\cS(\cA \otimes \cB) \supseteq,..., \supseteq \cS_{\beta}
\supseteq, ..., \supseteq conv(\cS(\cA)\otimes\cS(\cB)).$$
Here, $\C_{inj}$ stands for the injective cone, $\C_{\beta}$ for a tensor cone,
while $\C_{pro}$ for the projective cone. The tensor cone $\C_{\beta}$ 
is defined by the property: the canonical bilinear mappings 
$\omega :\cA_h \times \cB_h
\to (\cA_h \otimes \cB_h, \C_{\beta})$ and $\omega^* : \cA^*_h \times \cB^*_h \to 
(\cA^*_h \otimes \cB^*_h, \C_{\beta}^*)$ are positive. 
The connes $\C_{inj}, C_{\beta}, C_{pro}$ are different unless either
$\cA$, or $\cB$, or both $\cA$ and $\cB$ are abelian. 
This feature is the origin of various positivity concepts
for non-commutative composed systems and
it was Stinespring who used the partial transposition 
(transposition tensored with identity map) for showing the difference
among $C_{\beta}$ and $\C_{inj}$ and $\C_{pro}$. Clearly, in dual terms, the mentioned property
corresponds to the fact that separable states are different from the set of all states
and that there are various special subsets of states if both subsystems are truly quantum.

To summarize one can say that contrary to the plain positivity,
 CP property plays a dominant role 
in the programme of quantization of classical results for composed systems. 
However, as it will be discussed in the final section, other types of positivity
are helpful for better understanding the relations between various subsets of
states, the algebraic and the order structure.

\section{QUANTUM CORRELATIONS}

Now we wish to discuss problems associated with the partial order structure
of tensor product of \Cs-algebras which are related to quantum information theory.
Our first remark is the observation that  quantum information
theory relies on the fact that the restriction of a pure state of a composed system
to a subsystem is, in general, not pure. Moreover, if the restriction of a pure 
state is pure then the state of the composed system is of the product form.
This leads to the observation that the
coupling of observables of a composed system by an entangled state
 offers additional possibilities
for information exchange as well as a chance to reproduce states.
All that follows from the fact that entangled states exhibit non-classical correlations.

To be more precise, let $\cA_1 \subseteq \cB(\cH_1)$ and $\cA_2 \subseteq \cB(\cH_2)$ 
be two concrete \Cs-algebras and
define, for a state $\omega$ on $\cA_1 \otimes \cA_2$, the following map:
$$(r_1 \omega)(A) \equiv \omega(A \otimes {\bf 1})
\quad ( (r_2\omega(B) = \omega( {\bf 1}\otimes B) )$$
where $A \in \cA_1$ ($B \in \cA_2$).
\smallskip
 $r_{1(2)} \omega$ is a state on $\cA_{1(2)}$ .
Moreover:
{\it Let $(r_1 \omega)$ be a pure state on $\cA_1$.
 Then $\omega$ can be written as a product state on
$\cA_1 \otimes \cA_2$.}

Let $\omega$ be a state on ${\mathcal A}_1 \otimes \cA_2$. 
{\it The entanglement of formation}, EoF, of $\omega$ can be
 defined
as ( \cite{Mjp}, see also \cite{Ben})
$$
{E}(\omega) = \inf_{\mu \in M_{\omega}(\cS)}
\int_{\cS} d\mu(\varphi) S(r\varphi)
$$
where $S(\cdot)$ stands for the von Neumann entropy, i.e. $S(\varphi)
= - Tr \varrho_{\varphi} \log \varrho_{\varphi}$
 where $\varrho_{\varphi}$
is the density matrix determining the state $\varphi$. We want to stress that
other entropy-functions 
can be used! The given definition of EoF 
is based on the decomposition theory and in particular
$M_{\omega}(\cS) \equiv \{ \mu: \omega = \int_{\cS}\nu d\mu(\nu)\}$.
We recall that the separable states are those which are in the closure of the convex hull of 
simple tensors (so tensor products of subsystem states) while 
an entangled state stands for a non-separable one. One can prove \cite{Mjp}

\begin{theorem}
A state $\omega \in \cS$ is separable if and only if $E(\omega)$
is equal to 0.
\end{theorem}
Let us denote the set of all states on $\cA \equiv \cA_1 \otimes \cA_2$ ($\cA_1$, $\cA_2$)
by $\cS(\cA)$ ($\cS(\cA_1)$, $\cS(\cA_2)$ respectively).
Obviously,
 $r_i \omega$ is in $\cS(\cA_i)$, $i=1,2$.
Next, take a measure $\mu$ on $\cS(\cA)$. Then, using the restriction maps $r_i$ one can define
measures $\mu_i$ on $\cS(\cA_i)$ in the following way: for a Borel subset
$F_i \subset \cS(\cA_i)$ we put
\begin{equation}
\mu_i(F_i) = \mu(r_i^{-1}(F_i)), \quad i =1,2.
\end{equation}
Having measures $\mu_1$ and $\mu_2$, both coming 
from the given measure $\mu$ on $\cS(\cA)$, one can define new measure $\boxtimes \mu$
on $\cS(\cA_1) \times \cS(\cA_2)$ which encodes classical correlations
between the two subsystems described by $\cA_1$ and $\cA_2$ respectively (see \cite{Ma2} for details).
The measure $\boxtimes \mu$ leads to the concept of coefficient of local (quantum) correlations
for $\phi \in \cS(\cA), a_1 \in \cA_1, a_2 \in \cA_2$, which is defined as
\begin{eqnarray}
d(\phi, a_1, a_2)& = & \inf_{\mu \in M_{\phi}(\cS(\cA))}
|\phi(a_1 \otimes a_2) \nonumber \\
&& - (\int \xi d(\boxtimes \mu)(\xi))(a_1 \otimes a_2)| \nonumber
\end{eqnarray}

The crucial property of the coefficient of quantum correlations is that  
$d(\phi, \cdot \cdot)$ is equal to $0$ if and only if
the state $\phi$ is separable (\cite{Ma2}, \cite{Ma3}). 
The advantage of using $d(\cdot)$ lies in the fact that
that concept looks more operational and that it does not use an entropy function.
Moreover, $d(\cdot)$ is nothing else but the ``quantization'' of the classical concept
of coefficient of independence. Hence, we got a strong indication that entangled states 
contain new type of correlations which are called quantum.

\section{SOME APPLICATIONS}

\subsection{QUANTUM STOCHASTIC DYNAMICS (\cite{MZ1}-\cite{MZ5}, \cite{MOZ})}

\smallskip
It is well known that in the theory of classical particle systems one of the 
basic objectives is to produce, describe 
and analyze dynamical systems
with an evolution originated from stochastic processes in such a way that their 
equilibrium states are given Gibbs states (cf. \cite{Ligget}). 
A well known illustration of such an approach are systems with the so 
called Glauber dynamics \cite{Rx}. To carry out the analysis of 
dynamical systems with evolution originated from stochastic processes, it is convenient 
to use the theory of Markov processes in
the framework of $L_p$-spaces. 
In particular, for the Markov-Feller processes, using the unique correspondence
between the process and the corresponding dynamical semigroup, one can give a recipe 
for the construction of Markov generators (see \cite{Ligget}).
The correspondence uses the concept of conditional expectation which can be nicely
characterized within the (classical) $L_p$-space framework (cf. \cite{Moya}). Furthermore, 
(classical) $L_p$ spaces 
are extremely useful in a detailed analysis of the ergodic properties of 
the evolution.

However, as contemporary science is based on {\it quantum mechanics}, 
it is again legitimate
to look for a quantization of the above approach.
That task was carried out 
in the setting of quantum mechanics and the main ingredient of the quantization 
was the concept of generalized 
conditional expectation and 
Dirichlet forms defined in terms of non-commutative (quantum) $L_p$-spaces.
We already met these spaces in the description of quantized Banach-Lamperti theorems.
The advantage of using quantum $L_p$-spaces for the quantization of stochastic
dynamics lies in the fact that we can 
follow the traditional ``route'' of analysis of dynamical systems and that 
it is possible to have a single scheme for the quantum counterparts of stochastic dynamics of
both jump and diffusive type.

Turning to concrete dynamical systems, for example to jump type evolutions, 
we recall that one of the essential ingredients of the $L_p$-space approach to the 
analysis of such evolutions, is the usage of  
local knowledge. To illustrate that idea let us 
consider a region $\Lambda_I$ (usually finite) and its environment
 $\Lambda_{II}$. Then, performing
an operation over $\Lambda_I$ (e.g. a block-spin flip or a symmetry transformation) 
one is changing locally the reference
state. Such a change can be expressed in terms of generalized conditional expectations. 
Guided by the classical theory, one
can define, now in terms of generalized conditional expectations, 
the infinitesimal generator of quantum dynamics. It is
important to note that such a dynamics is the result of local operations 
(associated with the mentioned local knowledge about the system). \\
Then, having defined the dynamics, we should pose the natural 
question of its nontriviality. By this we understand,
first of all, that the infinitesimal generator of the dynamics is {\em not} a 
function of the hamiltonian defining the reference Gibbs
state. This requirement arises in a natural way from the methodology 
of constructing stochastic dynamics as sketched
in the preceding paragraph. In fact, it has been shown \cite{MOZ} that 
generators defined within the $L_p$-space
setting satisfy the above requirement. On the other hand, 
{\it in order to confirm that the constructed dynamics are interesting,
and the genuine quantum counterparts of classical dynamical maps it is necessary 
to study the evolution of entanglement and
correlations as measures of coupling between 
two subsystems} caused by local
(e.g. block-spin flip) operations.
Going in that direction, an analysis of stochastic quantum models based
on reference systems determined by Ising type and XXZ 
hamiltonians (\cite{KM2}, see also \cite{KM1}) was done. It has 
shown the tendency of enhancement of quantum correlations.
In the first example, based on one dimensional Ising model with nearest neighbor
interactions,  the lack of production of quantum correlations was shown.
This is to be expected because the Ising model illustrates a behaviour
typical of classical interactions (cf \cite{BR}).
The second example, based on the quantum XXZ model with more interesting
and complicated features of propagation, provides  {\it clear 
signatures of production of quantum correlations.}

\subsection{POSITIVE MAPS VERSUS ENTANGLEMENT}

The analysis of evolution of entanglement which was described in the previous subsection
indicates that there is a need for an operational measure of entanglement. 
This demand is strenghtened by the observation that the amount of states that can be used for
quantum information is measured by the entanglement.
On the other hand, the programme of classification of entanglement seems to be 
a very difficult task.
In particular, it was realized that the first step must presumably take
the full classification of all positive maps. To see this let us  take
a positive map $\alpha_{1,t}: \cA_1 \to \cA_1$, $t$ being the time, and consider
the evolution of a density matrix $\varrho$ 
($\varrho$ determines the state $\phi \in \cS(\cA \otimes \cB)$), 
i.e. we wish to study  $(\alpha_{1,t} \otimes id_2)^d\varrho$.
Here $(\alpha_{1,t} \otimes id_2)^d$ stands for the dual map, i.e. for the dynamical map in
the Schr\"odinger picture. 
Then, if $\varrho$ is an entangled state,
$(\alpha_{1,t} \otimes id_2)^d\varrho$ may develop negative eigenvalues and thus lose consistency as 
a physical state.
That observation was the origin of rediscovery, now in the physical context, of Stinespring's argument
saying that the tensor product of transposition with the identity map
can distinguish various cones in the tensor product structure (see Section 2).
This led to the criterion of separability (\cite{P}, \cite{H}) saying that only separable states
are globally invariant with respect to the familly of all positive maps.
However, criterions of that type are not operational. Even worse, they are strongly
related to a classification of positive maps.
In particular, the old open problem concerning the description of non-decomposable maps
was revived. To describe that problem we need (cf \cite{St}):

Let
$\tau : \cA \to \cB(\cH)$ be a linear, positive map. $\tau$ is called decomposable 
if there exists a
Hilbert space $\cK$, a bounded linear map $V: \cH \to \cK$ and a \Cs-homomorphism
$\pi : \cA \to \cB(\cK)$ such that $\tau = V^* \pi V$. 
\Cs-homomorphism means that $\pi (\{a,b\}) = \{\pi(a), \pi(b)\}$ where
$\{ \cdot, \cdot \}$ stands for anticommutator, i.e. $\pi$ preserves the Jordan structure!
A more subtle notion is the following:
$\tau$ is locally decomposable
if for $0 \ne x \in \cH$, there exists a Hilbert space $\cK_x$, $V_x:\cK_x \to \cH$
and a \Cs-hommomorphism $\pi_x$ of $\cA$ to $\cB(\cK_x)$ such that  
$$V_x \pi_x(a) V^*_x x = \tau (a)x$$
for all $a \in \cA$. 

\smallskip
It is known (\cite{Wor}, \cite{Choi}) that for the case $M_k(\bC) \otimes M_l(\bC)$
with $k=2=l$ and $k=2$, $l=3$ all positive maps are decomposable.
Then, the criterion for separability simplifies significantly. Namely,
to verify separability it is enough to analyse $(\tau \otimes id)^d$, 
with $\tau$ being the transposition, as
other positive maps are just convex combinations of CP maps (they always 
map states into states) and the composition of CP map with $\tau \otimes id$.

The situation changes dramatically when both $k$ and $l$ are larger than 2. 
In that case there are plenty of non-decomposable maps 
(see \cite{Kos} and the references given there) and to analyse entanglement
one cannot restrict oneself to study $\tau \otimes id$.
Thus, a full description of positive maps is needed.
Furthermore, one can constuct examples of entangled states using concrete 
non-decomposable maps (see \cite{HKP}). However, the classification of non-decomposable
maps is a difficult task which is still not completed (\cite{St3}, \cite{LMM}).

We want to close the section with an important remark. Namely, if $d(\phi, A)= 0$
for any $A \in \cA_1 \otimes \cA_2$ then, using the description of locally
decomposable maps, one can show that the state $\phi$ is separable  \cite{Ma3}.
This result shows how strong the interplay between separability and certain subtle
features of positive maps is. However, this is not unexpected as the full correspondence
between Schr\"odinger and Heisenberg picture relies on the underlying algebraic structure and
geometry of the state space, see \cite{A}, \cite{C}, and \cite{E}.

\section{Acknowledgements}
The author would like to thank the organisers of the XXXV Symposium on Mathematical
Physics, Torun, Poland for a very nice and interesting conference where the main topics of
this paper were presented. Thanks also to Mark Fannes for careful reading the manuscript.
The support of BW grant
5400-5-0255-3is gratefully acknowledged.

\end{document}